# Tuning the domain wall orientation in thin magnetic strips by induced anisotropy

S. Cherifi[*]

Institut Néel, CNRS, BP166, F-38042 Grenoble, France

R. Hertel

Institut für Festkörperforschung, Forschungszentrum Jülich GmbH, IFF-9, D-52425 Jülich, Germany

A. Locatelli

Elettra - Sincrotrone Trieste, S.S. 14, km 163.5, 34012 Basovizza (TS), Italy

Y. Watanabe

Japan Synchrotron Radiation Research Institute SPring-8, 1-1-1 Kouto Sayo-cho Sayo-gun Hyogo 679-5198 Japan

G. Potdevin[†], A. Ballestrazzi[‡], M. Balboni[‡] and S. Heun[§]

CNR-INFM Laboratorio Nazionale TASC, Area Science Park S.S. 14, km 163.5, 34012 Basovizza (TS), Italy

*Abstract*

We report on a method to tune the orientation of in-plane magnetic domains and domain walls in thin ferromagnetic strips by manipulating the magnetic anisotropy. Uniaxial in-plane anisotropy is induced in a controlled way by oblique evaporation of magnetic thin strips. A direct correlation between the magnetization direction and the domain wall orientation is found experimentally and confirmed by micromagnetic simulations. The domain walls in the strips are always oriented along the oblique evaporation-induced easy axis, in spite of the shape anisotropy. The controlled manipulation of domain wall orientations could open new possibilities for novel devices based on domain-wall propagation.

---

[*] E-mail address: salia.cherifi@grenoble.cnrs.fr
[†] Also at Institut Néel, CNRS, BP166, F-38042 Grenoble, France
[‡] Also at Dipartimento di Fisica, Universita' di Modena e Reggio Emilia, Via Campi 213/A, 41100 Modena, Italy
[§] Also at NEST CNR-INFM Scuola Normale Superiore, Piazza dei Cavalieri 7, 56126 Pisa, Italy

In soft-magnetic thin strips, the magnetization naturally aligns with the strip axis to minimize the magnetostatic energy. If the strips are sufficiently long, the magnetic structure may split up into domains with opposite magnetization. The boundary regions between ferromagnetic domains of opposite direction are known as domain walls (DWs). Recently, DWs in thin strips have been proposed as carriers of information, which can be transmitted either by external magnetic fields [1] or electric current pulses [2]. This kind of DW displacement is used in novel concepts of devices with a magnetic logic architecture referred to as "domain-wall logic" [3, 4] and in a "racetrack" design based on current-induced domain wall propagation [5]. In these devices, the information is transported and elaborated as the DWs propagate along a complex network of magnetic thin strips. For these reasons domain walls are nowadays considered as magnetic structures with promising technological potential.

Typical DWs occurring in magnetic strips are the so-called *head-to-head* type domain walls, which have been studied thoroughly in the last years [6, 7, 8, 9]. Most of these studies focused on establishing the phase diagram of the DW structures, namely "vortex" or "transverse" walls, as a function of two extrinsic parameters: the magnetic film thickness and the nanostructure widths. Other extrinsic parameters such as topographic defects or edge roughness were found to influence the DWs propagation velocity, as well [10]. Besides the extrinsic parameters, other parameters linked to the material's intrinsic properties influence the DW characteristics. With intrinsic properties, we intend properties such as saturation magnetization and spin polarization of the conduction electrons [11] or magnetic anisotropy and magnetization direction. Only little is known about the effect of these intrinsic properties on the DW spin structure and on the spin torque effect and its related DW propagation. The control of these internal parameters is a promising approach for the optimization of the DW propagation velocities and thus, for the previously mentioned applications.

We report on an approach for obtaining a tuneable orientation of in-plane magnetic domains and domain walls in thin magnetic strips. This is achieved by acting on the magnetic anisotropy; one of the material intrinsic properties. In particular, we show a direct influence of a uniaxial magnetic anisotropy on the DW orientation in thin Co strips. The direction of the in-plane uniaxial anisotropy is controlled by oblique evaporation (OE) along different azimuth angles. The oblique incident deposition is known to induce, via the so-called self-shadowing effect [12], anisotropic film growth that shows crystallographic texture. This anisotropic film texture is directly connected to the magnetocrystalline anisotropy. In obliquely evaporated films, both the shape anisotropy (morphology) and the magnetocrystalline anisotropy (texture) contribute to the uniaxial magnetic anisotropy. The easy magnetization axis is oriented in the film plane and can be either parallel or perpendicular to the evaporation direction, depending on the evaporation angle. When the evaporation angle is below 75° (with respect to the surface normal), the in-plane easy magnetization axis is oriented perpendicular to the evaporation direction, while it is parallel for angles above 75° [13]. While OE has long been known to generate a uniaxial magnetic anisotropy [14], the effect of the anisotropy on the domain wall orientation and its spin structure has not been reported yet.

In this experiment, 5-8 nm thick polycrystalline Co films are evaporated at grazing incidence on pre-patterned Si substrates, on which submicron-strips have been designed by electron-beam lithography (Fig. 1). A polar evaporation angle of 73° ensures that the in-plane magnetization axis is perpendicular to the evaporation direction. The effect of the OE-induced in-plane magnetic anisotropy on the DW spin structure in thin magnetic strips is investigated by high lateral resolution x-ray magnetic imaging in combination with micromagnetic simulations performed with a finite element-algorithm see, e.g. [15].

High resolution photoemission electron microscopy in combination with x-ray magnetic circular dichroism (XMCD-PEEM) has been carried out at the synchrotron Elettra (Trieste, Italy) using a commercial version of the Spectroscopic-LEEM microscope [16]. The samples have been grown in-situ, and the magnetic domain configurations have been analyzed in the virgin state (as-grown state) by tuning the energy of circularly polarized X-rays to the Co $L_3$ edge at two opposite elliptical helicities. The maximum contrast is obtained when the easy magnetization axis is oriented parallel or antiparallel to the direction of the incident photons. Therefore, the cobalt evaporator (flux direction) has been oriented perpendicular to the direction of the incoming X-rays. In order to compare different configurations, strips with three different azimuth orientations have been patterned on the same Si substrate, as shown in figure 1.

As a reference system, continuous cobalt films have been deposited under identical conditions and various azimutal angles on flat Si(111) (*i.e.* unpatterned surfaces). Figure 2 (a) shows hysteresis loops measured on such a 8 nm thick polycrystalline Co film by longitudinal magneto-optical Kerr. The external magnetic field is applied in the plane, along three different azimuth angles with respect to the evaporation direction (cf. inset of Fig. 2 (a)). The easy magnetization axis is always found to be in the surface plane and perpendicular to the evaporation direction, independent of the in-plane crystallographic orientation of the substrate. The measured in-plane coercive field is approximately 200 Oe, which is around ten times larger than the expected value for a film deposited at normal incidence (13 – 26 Oe) [17]. The saturation field along the hard axis is about 500 Oe. The polar representation (not shown) of the magnetization intensity along different azimuth angles shows the uniaxial character of the OE-induced anisotropy and confirms that the easy magnetization axis is perpendicular to the Co flux. Elongated magnetic domains of alternating opposite magnetization perpendicular the evaporation flux direction are observed by XMCD-PEEM, as shown in figure 2 (b). These 3-4 µm wide magnetic domains are separated by 180° Néel walls and are aligned along the OE-induced easy magnetization axis.

In the case of Co strips, different micromagnetic configurations are expected because of the competition between the OE-induced anisotropy and the shape anisotropy. Therefore, we investigated three different geometries where the OE-induced easy magnetization axis is oriented parallel, perpendicular, and at 45° with respect to the axis of the strips.

When the OE-induced easy axis is parallel to the axis of the strips (Fig. 3 (a)), 0.5 µm wide Co strips are found to be in a single-domain magnetic configuration. In this case, the magnetization orients along the strips, as imposed by both shape and OE-induced anisotropies. Our micromagnetic simulations confirm this observation and show that if a domain wall is present, it is aligned along the strips.

Figure 3 (b) shows the magnetic domain structure obtained when the easy magnetization axis is oriented perpendicular to the strips. In this case, a multi-domain configuration is observed where 180º DW separating two opposite domains are along the strips width. Therefore in this case the induced anisotropy is stronger than the shape anisotropy, since the latter tends to align the magnetization along the strips. This result is observed for different strip widths ranging from 0.5 µm to 2 µm, and is in agreement with micromagnetic simulations, which yield 180° DWs oriented perpendicular to the strips, cf. Fig. 3 (c).

A multi-domain magnetic configuration is also observed in submicron-strips when the OE-induced magnetization axis is at an intermediate angle between the two previous geometries (Fig. 4 (a)). In this case, the magnetic domains are separated by inclined 180º DWs. The orientation of these DWs mostly coincides with the OE-induced magnetic easy axis. The DWs contain a vortex spin structure, as evidenced by the micromagnetic simulations. Inclined domain walls in ultra-thin epitaxial Fe strips have also been reported by Vedmedenko *et al*. [18]. In that study, it was found that the orientation of the DW follows crystallographic directions in spite of shape anisotropy. This effect has been attributed to the presence of an anisotropic exchange interaction. Here, we show the presence of a clear correlation between the magnetic anisotropy and the domain wall orientation.

The inclined orientation of the DWs in thin strips may appear as a surprising result. From the point of view of the exchange energy, it would be more advantageous to have DWs oriented perpendicular to the strip, thereby reducing the length of the DWs and thus the total exchange energy. Since the magnetization direction changes by 180° for any orientation of the DWs, the exchange energy density does not depend on the DW orientation. We thus attribute the alignment of the DWs with the easy axis to magnetostatic effects: Only if the DW is parallel to the magnetization direction in the domains, an equal amount of positive and negative magnetic volume charges $\rho = -\nabla M$ is formed in the DW. These charges compensate each other and therefore this DW orientation represents an optimal DW structure from a magnetostatic point of view. Any other DW orientation would lead to charged domain wall types, which would lead to an increase of the magnetostatic field and the dipolar energy. Hence, the orientation of 180° DWs in thin strips has a tendency to align with the magnetization direction in the adjacent domain.

When an intermediate anisotropy angle (about 45°) is induced in wider strips of 1-2.5 µm, a more complex DW structure is observed, as shown in figure 4 (b). A gradual rotation of the DWs from 45° –in the centre of the microstructure– to parallel to the strips –close to the borders– is evidenced both experimentally and by micromagnetic simulations. In particular, inclined DWs with an S-like shape are evidenced in both simulations and XMCD-PEEM images. We attribute these to a non-uniform result of the competition between the OE-induced anisotropy and shape anisotropy. The latter is stronger in regions close to the borders and tends to rotate the magnetization along the axis of the strips, while the OE-induced anisotropy dominates in the centre of wide strips. The gradually different orientation of the magnetization from the center of the strip to the boundary and the observation of a curved, S-shaped DW is consistent with the previous argument concerning the correlation between the DW orientation and the magnetization direction in the domains.

We have also performed the same set of measurements on Fe instead of Co using substrates with either (001) or (111) orientation (not shown here). Our results indicate that the effect of the anisotropy on the DW orientation is not specific to Co strips or to the orientation of the substrate, and that it can be generalized to other materials and substrates. We thus conclude that oblique evaporation is a suitable method to control domain walls in thin magnetic strips. Its generalization to other materials such as magnetic alloys and diluted magnetic semiconductors opens new possibilities for better understanding the interaction between DWs and electrical currents, in particular in the case of inclined DWs [19].

In conclusion, we have shown a simple way to control the magnetic anisotropy and DW orientation in thin magnetic strips. We have induced a uniaxial in-plane magnetic anisotropy by oblique evaporation. This anisotropy is found to dominate in thin nanostructures, and it has been exploited for controlling DW orientations. The control of the DW orientation in thin strips is expected to open new possibilities for the ongoing research on how to exploit DWs as rapid and versatile carriers of information, by altering their static and presumably also their dynamic properties.

S.C. thanks M. Bonfim for help with Kerr measurements.

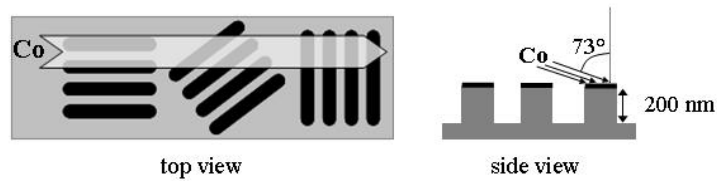

**Fig.1:** Schematics of the sample geometry where the Co flux is evaporated at oblique incidence on a pre-patterned substrate

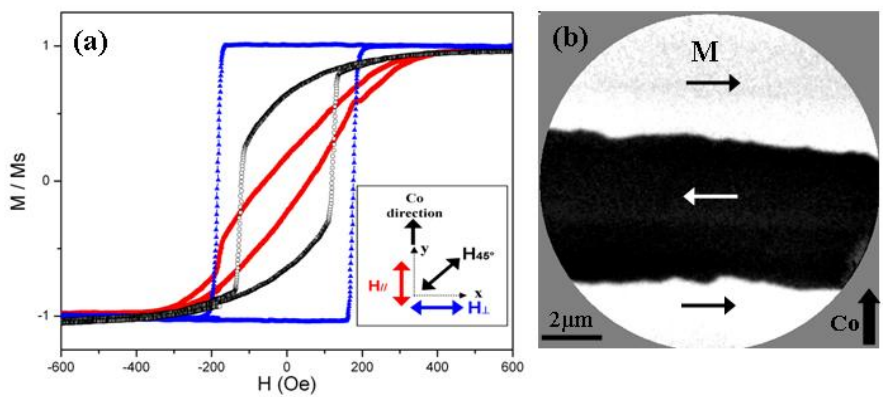

**Fig. 2:** (**a**) The hysteresis loops measured longitudinally in an 8 nm thick Co/Si(111) continuous film evaporated at oblique incidence show an easy magnetization axis oriented perpendicular to the Co flux direction (image inset). (**b**) The corresponding XMCD-PEEM image measured in the virgin state shows large magnetic domains along the easy magnetization axis.

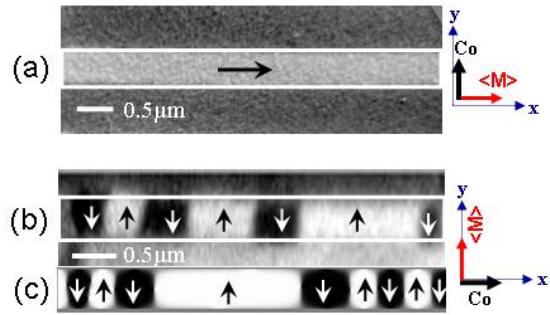

Fig. 3: Magnetic domain configurations observed by XMCD-PEEM in 5 nm thick Co nanostrips where the easy magnetization axis is: (a) parallel to the strips' axis; (b) perpendicular to the strips' axis. The magnetic domain configuration is confirmed by micromagnetic simulations (c).

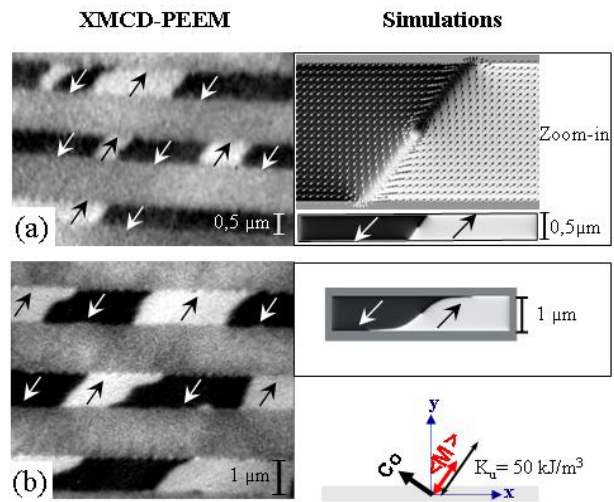

Fig.4: XMCD-PEEM images (left) and the corresponding micromagnetic simulations (right) showing inclined domain walls in 5 nm thick Co (a) nanostrips and (b) microstrips where a uniaxial magnetic anisotropy is induced at 45° with respect to the strips' axis, as indicated by arrows in the sketch